\documentclass[fleqn,usenatbib,useAMS]{mnras}

\usepackage{newtxtext,newtxmath}

\usepackage[T1]{fontenc}

\DeclareRobustCommand{\VAN}[3]{#2}
\let\VANthebibliography\thebibliography
\def\thebibliography{\DeclareRobustCommand{\VAN}[3]{##3}\VANthebibliography}

\usepackage{graphicx}	
\usepackage{amsmath}	
\usepackage{orcidlink}

\newcommand{\msun}{{\rm M}_{\sun}}

\title[Magnetic double-faced white dwarfs]{Double-faced white dwarfs and the magnetic inhibition of convection}

\author[S. Ginzburg]{
Sivan Ginzburg$^{\orcidlink{0000-0002-3751-4553}}$\thanks{E-mail: \href{mailto:sivan.ginzburg@mail.huji.ac.il}{sivan.ginzburg@mail.huji.ac.il}}
\\
Racah Institute of Physics, The Hebrew University, Jerusalem 9190401, Israel
}

\date{Accepted XXX. Received YYY; in original form ZZZ}

\pubyear{\the\year{}}

\begin{document}
\label{firstpage}
\pagerange{\pageref{firstpage}--\pageref{lastpage}}
\maketitle

\begin{abstract}
About one in five white dwarfs undergoes spectral evolution from a helium atmosphere to hydrogen and then back to helium. These short-lived hydrogen envelopes -- the result of residual hydrogen diffusion -- are eventually destroyed by either hydrogen or helium convection. An emerging class of double-faced white dwarfs seems to catch this process in the act, with varying amounts of hydrogen across regions of the stellar surface. Here, we quantitatively test the hypothesis that these inhomogeneities are the result of the magnetic inhibition of convection. We compute the critical magnetic field $B_{\rm crit}(M,T_{\rm eff})$ required to inhibit convection in both hydrogen and helium for $0.6-1.2\,\msun$ white dwarfs using two methods. Initially, we estimated $B_{\rm crit}\sim\sqrt{8\upi P}$ where $P$ is the pressure at the base of the convection zone, finding that most (three out of four) of the observed magnetic double-faced white dwarfs could potentially be explained by the magnetic inhibition of hydrogen convective energy transfer, with measured $B\gtrsim B_{\rm crit}^{\rm H}$. Then, we incorporated the magnetic field consistently into the stellar structure and directly computed the boundary of convective mixing. With this more appropriate method, we find that only half (two out of four) of the stars could be explained by the magnetic inhibition of helium convection, with $B\gtrsim B_{\rm crit}^{\rm He}$. Specifically, order of unity variations in the magnetic field's strength or orientation across the surface could account for the double-faced nature of these stars. Given our mixed results, other -- including non-magnetic -- scenarios should be considered as well.
\end{abstract}




\section{Introduction}

Observations of white dwarfs with different effective temperatures (and therefore different ages) reveal that about 70 (10) per cent of white dwarfs retain hydrogen (helium) atmospheres throughout their cooling. The remaining 20 per cent undergo spectral evolution as they cool down: their outermost atmospheres transition from helium to hydrogen and then back to helium \citep{Bedard2024,Kilic2025}.
The first transition is explained by the float-up of residual hydrogen from the interior under the influence of the white dwarf's strong gravity, forming a thin pure-hydrogen envelope. Later on, this thin layer is destroyed by convection, explaining the second spectral transition. Specifically, if the superficial hydrogen layer is lighter than $\sim 10^{-14}\,\msun$, it is eroded by convection in the underlying helium layer (this process is referred to as convective dilution). A heavier hydrogen layer shields the underlying helium from convection, but becomes convective itself at a lower effective temperature (this case is referred to as convective mixing). For more details, see \citet{FontaineWesemael1987,MacDonaldVennes1991,Rolland2018,Rolland2020,Bedard2023,Bedard2024}.

Much of our empirical understanding of white dwarf spectral evolution comes from measuring the ratio of spectral types as a function of effective temperature for large white dwarf samples \citep{Bedard2024,Kilic2025}. The emerging class of double-faced white dwarfs \citep{Caiazzo2023,Moss2024,Moss2025,BedardTremblay2025,Cheng2025} may offer  a complementary -- more direct -- route to probe spectral evolution.\footnote{\citet{Caiazzo2023} sparked the current interest in double-faced white dwarfs, though a few other members of this class were identified decades ago \citep[e.g.][]{Achilleos1992,Pereira2005}.} These white dwarfs exhibit periodic variations in the strength of their hydrogen and helium lines, presumably modulated by the white dwarf's rotation. The most extreme specimen of the class \citep{Caiazzo2023} oscillates between a pure DA (hydrogen) and a pure DB (helium) spectrum, although a very light ($\sim10^{-19}\,\msun$) hydrogen envelope with varying thickness may actually be covering the entire surface \citep{BedardTremblay2025}. In any case, double-faced white dwarfs seem to be characterized by an inhomogeneous distribution of hydrogen across their surface. As emphasized by \citet{BedardTremblay2025}, the hydrogen abundance (or thickness) does not necessarily vary between two hemispheres (as their name suggests); double-faced white dwarfs may instead have a different geometry, such as spots or polar caps and an equatorial belt with different compositions \citep{Beauchamp1993,Koester1994,Pereira2005,Moss2024,Moss2025}.    

Magnetic fields have been invoked early on as a possible explantation for the inhomogeneous surface distribution of double-faced white dwarfs \citep{Achilleos1992,Kidder1992}. Magnetic fields can suppress convective energy transfer \citep[e.g.][]{Tremblay2015}, such that a difference in the magnetic field's strength or orientation across the surface may lead to an inhomogeneous hydrogen envelope in white dwarfs that are currently undergoing convective mixing or dilution \citep[see also][]{Caiazzo2023}. This hypothesis is supported by the fact that about half of the double-faced white dwarfs are confirmed to be magnetic \citep{Moss2025}, compared to only 1 per cent of DB white dwarfs in the \citet{Genest-BeaulieuBergeron2019} sample. However, \citet{BedardTremblay2025} also discuss alternative scenarios for double-faced white dwarf formation, some of which do not require a magnetic field.

In this paper, we use the growing number of magnetic double-faced white dwarfs to quantify the effect of magnetism on white dwarf spectral evolution and to identify the origin of this intriguing class. Specifically, in Section \ref{sec:Bcrit} we compare the measured magnetic fields to the critical magnetic field required to inhibit convective energy transfer as a function of the white dwarf's mass and effective temperature, using a simple post-processing procedure. 
In Section \ref{sec:criterion}, we discuss the caveats in this simplistic approach, and compare the observations to a more accurate implementation of the criterion for convective instability. We summarize our conclusions in Section \ref{sec:conclusions}.  

\section{Critical magnetic field}\label{sec:Bcrit}

\citet{Moss2025} noted that the relatively high effective temperatures of the currently observed double-faced white dwarfs are consistent with helium convection (i.e. convective dilution). However, they also predicted the existence of cooler double-faced whited dwarfs as a result of hydrogen convection (convective mixing). To account for both options, we evolved white dwarfs with both hydrogen and helium envelopes using the stellar evolution code \textsc{mesa} \citep{Paxton2011,Paxton2013,Paxton2015,Paxton2018,Paxton2019,Jermyn2023}, version r23.05.1.

Specifically, we created carbon--oxygen (CO) core white dwarfs with masses $M=0.6$, $0.8$, and $1.0\,\msun$ using the \texttt{make\_co\_wd} test suite, which evolves progenitor stars with different initial masses from the pre-main-sequence through the various stages of stellar evolution. Similarly, we created a $1.2\,\msun$ oxygen--neon (ONe) core white dwarf using the \texttt{make\_o\_ne\_wd} test suite, but with the larger \texttt{sagb\_NeNa\_MgAl.net} nuclear reaction network instead of the original \texttt{co\_burn\_plus.net}, in order to produce a more accurate oxygen to neon ratio \citep[see][]{DeGeronimo2022}.\footnote{The core composition (i.e. CO or ONe) of ultramassive white dwarfs with $M\gtrsim 1.05\,\msun$ is still an open question (e.g. \citealt{Camisassa2022,BlatmanGinzburg2024} and references therein). None the less, we find that for the purposes of this study, our $1.2\,\msun$ ONe white dwarf behaves qualitatively similarly to the lower-mass CO white dwarfs.}
The white dwarf cooling stage was computed using the \texttt{wd\_cool\_0.6M} test suite, where we switched between hydrogen and helium envelopes using the \texttt{replace\_initial\_element} procedure. We emphasize that we do not assume a specific initial thickness for the superficial hydrogen layer (which is later mixed or diluted). Instead, we compute the emergence of convection in a pure hydrogen envelope (in case it is $\gtrsim 10^{-14}\,\msun$) or a pure helium envelope (in case the overlying hydrogen is $\lesssim 10^{-14}\,\msun$), and discuss the implications on the critical magnetic field below. 

\begin{figure}
\includegraphics[width=\columnwidth]{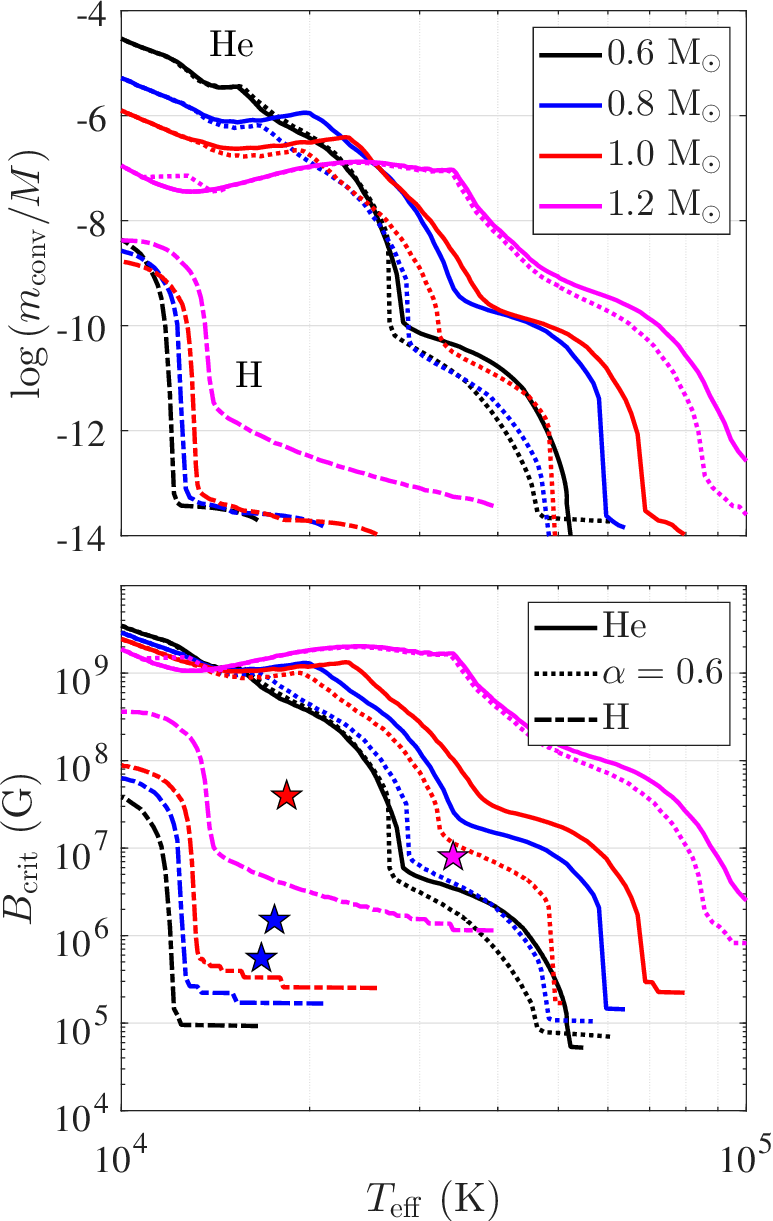}
\caption{\textit{Top panel:} the mass of the surface convection zone $m_{\rm conv}$ as a function of the white dwarf's effective temperature $T_{\rm eff}$ for several white dwarf masses $M$, and for both hydrogen (H, dot--dashed lines) and helium (He, solid lines) envelopes. We adopt a nominal mixing length $\alpha=1.8$, and also plot $\alpha=0.6$ for the helium envelopes (dotted lines; hydrogen envelopes are less sensitive for the purposes of this study). \textit{Bottom panel:} the critical magnetic field $B_{\rm crit}=\sqrt{8\upi P}$ required to suppress convective energy transfer, where $P$ is the pressure at the base of the convection zone. Star markers indicate the magnetic double-faced white dwarfs from \citet{Moss2025}, coloured as the model with the closest mass.}
\label{fig:EnvHeH}
\end{figure}

In the top panel of Fig. \ref{fig:EnvHeH}, we plot the mass of the convection zone $m_{\rm conv}$ as a function of the white dwarf's effective temperature $T_{\rm eff}$ for both hydrogen and helium envelopes. The formation of the convection zone and its growth as the white dwarf cools down are largely similar to \citet{Bedard2024}. Specifically, helium convection starts earlier due to its higher recombination temperature, and in both cases the convection zone initially remains very shallow before it rapidly expands inward. We used the ML2 mixing-length version \citep{BohmCassinelli1971,Tassoul1990} with a parameter $0.6\leq\alpha\leq 1.8$, which roughly bounds the calibration from three-dimensional simulations \citep{Tremblay2015calib,Cukanovaite2019}.

According to the standard theory of spectral evolution (see \citealt{Bedard2024} for a review), superficial hydrogen envelopes that are lighter than $\sim 10^{-14}\,\msun$ are eroded from below as convection develops in the underlying helium layer. On the other hand, three-dimensional simulations suggest that magnetic fields $B$ suppress convective energy transfer if the magnetic energy density exceeds the gas pressure $P$ \citep{Tremblay2015}
\begin{equation}\label{eq:simple_criterion}
    \frac{B^2}{8\upi}\gtrsim P.
\end{equation}
Taken at face value, equation \eqref{eq:simple_criterion} suggests that convective dilution in magnetic white dwarfs may be at least partially hindered until the convection zone deepens sufficiently (but see Section \ref{sec:criterion}).
Using hydrostatic equilibrium, this happens when the mass of the convection zone satisfies
\begin{equation}\label{eq:Pcrit}
   P\approx\frac{GM m_{\rm conv}}{4\upi R^4}\sim\frac{B^2}{8\upi}, 
\end{equation}
where in practice we evaluate $P(T_{\rm eff})$ at the base of the convection zone with \textsc{mesa}.\footnote{We emphasize that convection (or the lack of it) does not affect the white dwarf's cooling rate until the convection zone couples to the degenerate interior, which occurs at a much lower $T_{\rm eff}$ \citep{Fontaine2001,Tremblay2015,Ginzburg2024}.} $R$ denotes the white dwarf's radius, and $G$ is the gravitational constant. We plot the critical magnetic field $B_{\rm crit}$ required to delay convective energy transfer until a given $T_{\rm eff}$ (which corresponds to $m_{\rm conv}$) is reached in the bottom panel of Fig. \ref{fig:EnvHeH}. We assume that once energetically efficient helium convection begins at pressures $P\gtrsim B^2/(8\upi)$, overshoot above the convection zone dilutes the hydrogen envelope (similar overshoot is invoked in the standard picture of convective dilution).  

Hydrogen envelopes that are heavier than $\sim 10^{-14}\,\msun$ stifle helium convection, but become convective themselves at a lower $T_{\rm eff}$. In this case too, magnetic fields delay the onset of convective energy transfer until the convection zone deepens sufficiently, according to equation \eqref{eq:Pcrit}. We plot in Fig. \ref{fig:EnvHeH} $B_{\rm crit}(T_{\rm eff})$ that is required to delay hydrogen convective energy transfer, and therefore potentially affect convective mixing with the underlying helium layers (though see Section \ref{sec:criterion}). We note that in this case, even if $B<B_{\rm crit}$, convective mixing does not necessarily commence: if the hydrogen envelope is much thicker than $m_{\rm conv}(T_{\rm eff})$, convection (and even overshoot) does not reach the underlying helium until $T_{\rm eff}$ decreases further and $m_{\rm conv}$ grows \citep{Cunningham2020}. In other words, magnetic inhibition of convective mixing may require some fine-tuning of the hydrogen envelope's mass.

\subsection{Comparison with observations}\label{sec:obs}

Finally, we compare in Fig. \ref{fig:EnvHeH} our theoretical $B_{\rm crit}(M,T_{\rm eff})$ to the sample of magnetic double-faced white dwarfs from \citet{Moss2025}. In our model, we identify the measured magnetic field $B$ with the critical $B_{\rm crit}$ required to inhibit convective energy transfer. We assume that $B$ varies by a factor of a few across the stellar surface (e.g. a dipole), such that in some solid angles convective energy transfer is inhibited ($B\gtrsim B_{\rm crit}$) while in others it is not  ($B\lesssim B_{\rm crit}$) -- potentially explaining the double-faced nature of the white dwarf. Alternatively, the different orientation of the magnetic field (i.e. radial or azimuthal) across the surface may be the dominant factor; in this case too we expect $B\sim B_{\rm crit}$ in double-faced white dwarfs. As emphasized in Section \ref{sec:criterion} \citep[see also][]{Tremblay2015,Cunningham2021}, the magnetic inhibition of convective energy transfer does not necessarily imply the inhibition of convective mixing. The hypothesis in the current section however is that some asymmetry in the efficiency of convective energy transfer or mixing might -- perhaps indirectly -- alter the distribution of surface hydrogen depending on whether $B$ is above or below $B_{\rm crit}$ \citep[this is one of the scenarios proposed by][]{BedardTremblay2025}.

As seen in Fig. \ref{fig:EnvHeH}, it is challenging to explain the observed double-faced white dwarfs with helium convection (solid lines), because its associated $B_{\rm crit}$ is orders of magnitude stronger than the measured $B$. \citet{Caiazzo2023} and \citet{Moss2025} noted that a much lower $B_{\rm crit}\sim 10^4$ G is sufficient to inhibit convective energy transfer at the \textit{photosphere} rather than entirely as in our model \citep[see also][]{Tremblay2015}, but in this case it is unclear how the hydrogen is diluted in some parts of the stellar surface for the measured $B\gtrsim 10^6$ G (in other words, it is unclear how $B\gg B_{\rm crit}$ can lead to a double-faced configuration). 

Inspection of Fig. \ref{fig:EnvHeH} reveals another possibility. Although the relatively high effective temperatures of the observed double-faced white dwarfs are usually associated with helium convection, all the magnetic white dwarfs in \citet{Moss2025} have atypically high masses $M>0.6\,\msun$, such that hydrogen convection may also be relevant, because $T_{\rm eff}$ at the onset of convection increases with the surface gravity $g\equiv GMR^{-2}$. We note that previous studies found a weaker $T_{\rm eff}(g)$ dependence (\citealt{BauerBildsten2019}; A. B\'edard and P.-E. Tremblay, private communication), such that the high $T_{\rm eff}$ of our $1.2\,\msun$ model (which is marginally consistent with the observations) may be sensitive to modelling details. As seen in Fig. \ref{fig:EnvHeH}, inhibition of energy transfer in the nascent hydrogen convection zone (dot--dashed lines) at such $T_{\rm eff}$ requires $B_{\rm crit}\sim 10^5-10^6$ G. Specifically, for three out of the four stars in the sample (notwithstanding the numerical uncertainty in our $1.2\,\msun$ model), this $B_{\rm crit}$ is smaller than the observed $B$ by less than an order of magnitude, which may explain the double-faced nature of these white dwarfs. In other words, it is possible that for large portions of the stellar surface $B\lesssim B_{\rm crit}$, or that the magnetic field's orientation favours convective energy transfer.   

\section{Convective instability criterion}\label{sec:criterion}

While the equipartition critical magnetic field approach of Section \ref{sec:Bcrit} (but applied to the photosphere, not to the base of the convection zone) has been suggested in previous studies \citep{Caiazzo2023,Moss2025}, it suffers from several shortcomings. First of all, it is not clear to which extent the mixing of hydrogen and helium is inhibited for $B>B_{\rm crit}$. \citet{Tremblay2015} found that convective velocities remain similar even when most of the convective energy transfer is suppressed. Moreover, in the context of accreted metals transport, \citet{Cunningham2021} found that reducing the mixing efficiency by orders of magnitude does not necessarily result in surface inhomogeneity. A second issue is that a reduction in the convective energy transfer changes the atmosphere's temperature profile \citep{Tremblay2015}, shifting the region of convective instability for which $B_{\rm crit}$ was estimated in the first place. For example, \citet{Ginzburg2024} demonstrated that in certain circumstances even weak magnetic fields $B\ll B_{\rm crit}=\sqrt{8\upi P}$ may significantly push the boundary of the convection zone. 

To address these shortcomings, in this section we take an alternative approach and adopt the \citet{GoughTayler1966} criterion for convective instability \citep[see also][]{MullanMacDonald2001,Ginzburg2024}:
\begin{equation}\label{eq:guogh}
        \nabla_{\rm rad}>\nabla_{\rm ad}+\frac{B^2}{B^2+4\upi\gamma P},
\end{equation}
where $\nabla\equiv {\rm d}\,\ln T/{\rm d}\,\ln P$ denotes the temperature gradient (radiative or adiabatic), and $\gamma\equiv 1/(1-\nabla_{\rm ad})$ is the adiabatic index. 
For $B=0$, equation \eqref{eq:guogh} reduces to the regular Schwarzschild criterion $\nabla_{\rm rad}>\nabla_{\rm ad}$, such that the actual temperature gradient is $\nabla=\min(\nabla_{\rm rad},\nabla_{\rm ad})$. 
For $B>0$, equation \eqref{eq:guogh} takes into account the reduction in convective energy transfer for pressures $P\lesssim B^2/(4\upi\gamma)$, in accordance (approximately) with equation \eqref{eq:simple_criterion}. However, according to equation \eqref{eq:guogh}, even higher pressures are still affected by the magnetic field, which slightly raises the effective adiabatic gradient. At the base of the convection zone $\nabla_{\rm rad}\approx\nabla_{\rm ad}$, such that even a slight modification to $\nabla_{\rm ad}$ can shift the convective boundary.\footnote{The lower pressure at the new convective boundary makes it even more susceptible to magnetic fields, such that surprisingly low $B\ll B_{\rm crit}$ can shift the convection zone dramatically.} 
To construct self-consistent temperature profiles, we implement the \citet{GoughTayler1966} criterion in \textsc{mesa} by using the \texttt{other\_mlt\_results} hook and setting the actual temperature gradient to 
\begin{equation}\label{eq:gough_mesa}
\nabla=\min\left(\nabla_{\rm rad},\nabla_{\rm ad}+\frac{B^2}{B^2+4\upi P}\right),    
\end{equation}
where we have omitted $\gamma\sim 1$ for simplicity. Now, with this modified Schwarzschild criterion, we can focus on the regions where the magnetic field shuts down convection entirely -- including mixing -- and not only the convective energy transfer.

\begin{figure}
\includegraphics[width=\columnwidth]{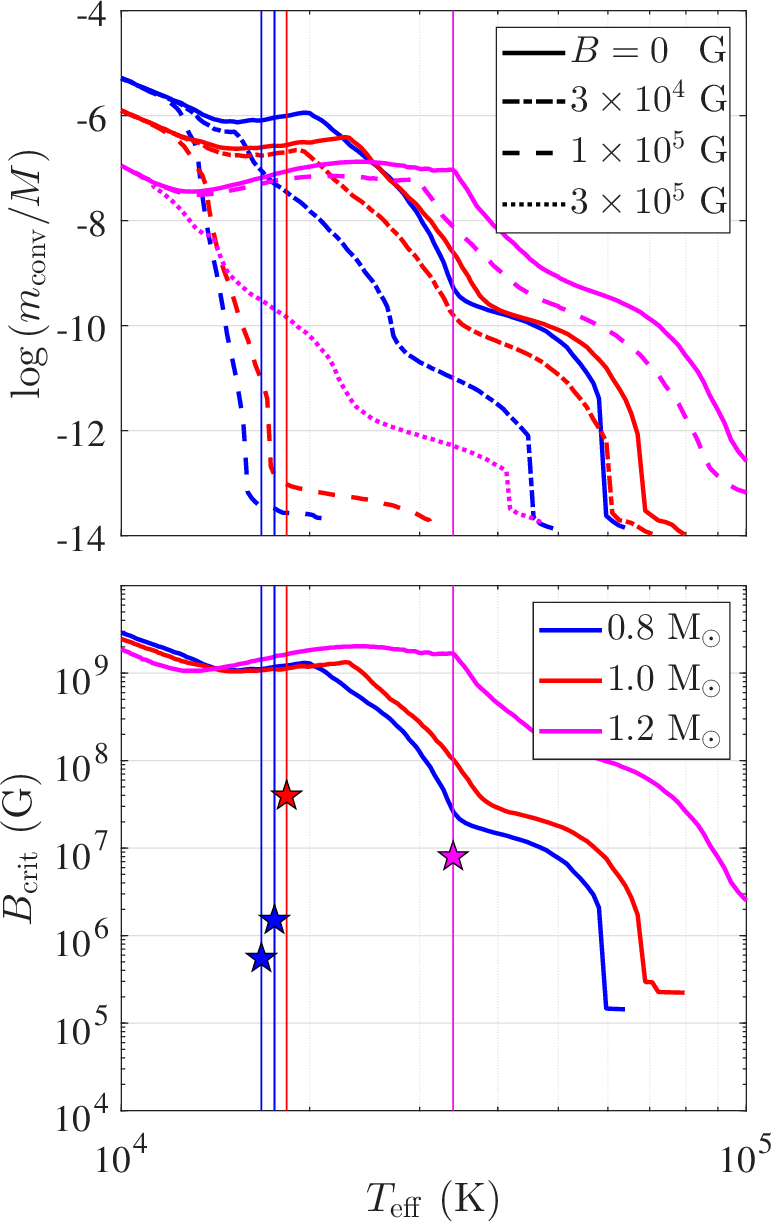}
\caption{\textit{Top panel:} the mass of the helium convection zone as a function of the magnetic field $B$, which is incorporated consistently into the stellar structure using the \citet{GoughTayler1966} prescription; see Section \ref{sec:criterion} (only some combinations of $B$ and $M$ are plotted to avoid clutter). \textit{Bottom panel:} for comparison, the critical magnetic field $B_{\rm crit}$ as computed by the simpler post-processing procedure of Section \ref{sec:Bcrit} (same as in Fig. \ref{fig:EnvHeH}).
The thin vertical lines (colour-coded by mass) mark the $T_{\rm eff}$ of the observed magnetic double-faced white dwarfs.}
\label{fig:Grad}
\end{figure}

As seen in Fig. \ref{fig:Grad}, the magnetic field $B$ -- when accounted for consistently -- pushes the convective boundary outwards to much lower pressures, even for $B\ll B_{\rm crit}$ (that is, the critical field as computed in Section \ref{sec:Bcrit}). Specifically, Fig. \ref{fig:Grad} (top panel) demonstrates that a \textit{revised} $B_{\rm crit}\sim 10^5$ G delays the growth of a significant helium convection zone until the observed $T_{\rm eff}$ is reached (thin vertical lines; this result is insensitive to $\alpha$). With measured $B\gg B_{\rm crit}\sim 10^5$ G, it is unclear how partial (i.e. on parts of the stellar surface) inhibition of convective dilution can explain the two massive double-faced white dwarfs in the sample. For the remaining two lower-mass ($\approx 0.8\,\msun$) double-faced white dwarfs, our revised $B_{\rm crit}$ is smaller than the observed $B$ by only a single order of magnitude, such that regions on the stellar surface with a weaker field or a favourable magnetic field orientation might provide a plausible explanation for the surface inhomogeneity. Inhibition of hydrogen convection requires even lower (revised) $B_{\rm crit}$ and is therefore not presented here.

We emphasize that although we implemented a very specific prescription for the magnetic inhibition of convection \citep{GoughTayler1966}, our results may actually be more general, at least qualitatively. Consistent magnetic mixing-length theories transition smoothly from strong to weak convection \citep{Stevenson1979,BessilaMathis2024,BessilaMathis2025}. Equation \eqref{eq:gough_mesa} may be thought of as a similar smoothed version of the \citet{Tremblay2015} criterion, taking into account the gradual inhibition of convective energy transfer as a function of $P$ (with $\sim 4\upi P/B^2$ as a natural smoothing parameter).

\section{Conclusions}\label{sec:conclusions}

The growing class of double-faced white dwarfs has been associated with the magnetic inhibition of convection, which controls the amount of hydrogen on the surface \citep{Achilleos1992,Caiazzo2023,Moss2025}. In this paper, we analysed this hypothesis quantitatively by comparing the measured magnetic fields $B$ of observed double-faced white dwarfs to the critical magnetic field $B_{\rm crit}(M,T_{\rm eff})$ required to suppress convection. We suggest that in order to explain the double-faced nature of the observations, the magnetic fields must satisfy $B\sim B_{\rm crit}$, such that convection is suppressed only in parts of the stellar surface due to order of unity variations in the magnetic field's strength or orientation across the surface.

We used \textsc{mesa} to compute the growth of the convection zone as the white dwarf cools down $m_{\rm conv}(T_{\rm eff})$ for both hydrogen and helium envelopes for white dwarf masses $M=0.6-1.2\,\msun$. Helium convection can destroy thin ($\lesssim 10^{-14}\,\msun$) overlying hydrogen atmospheres through convective dilution, whereas heavier hydrogen atmospheres stifle helium convection, become convective themselves at a lower $T_{\rm eff}$, and may eventually be destroyed through convective mixing with the underlying helium \citep{Bedard2024}. 

Initially, we estimated $B_{\rm crit}\sim\sqrt{8\upi P}\propto m_{\rm conv}^{1/2}$ by comparing the magnetic energy density to the gas pressure $P$ at the bottom of the convection zone \citep{Tremblay2015}. Based on the relatively high $T_{\rm eff}$ of the observed double-faced white dwarfs, \citet{Moss2025} focused on helium convection. However, we found that for all the magnetic white dwarfs in their sample $B\ll B_{\rm crit}^{\rm He}$ by several orders of magnitude, apparently ruling out this scenario (Fig. \ref{fig:EnvHeH}). Due to their atypically high masses $M>0.6\,\msun$, hydrogen convection is another possibility even at such high $T_{\rm eff}$. In fact, we found that for three out of the four stars, $B\gtrsim B_{\rm crit}^{\rm H}$ by less than an order of magnitude (Fig. \ref{fig:EnvHeH}), although the residual hydrogen convection zone of our $1.2\,\msun$ model at very high $T_{\rm eff}$ might be a numerical artifact. 

This post-processing (i.e. $m_{\rm conv}$ was computed assuming $B=0$) critical magnetic field calculation has several drawbacks. First, \citet{Tremblay2015} demonstrated that it is relevant for the magnetic inhibition of convective energy transfer, but not necessarily mixing. In addition, the reduced efficiency of energy transfer changes the temperature profile, moving the convection zone's boundary -- changing $B_{\rm crit}$. We addressed both of these issues by directly implementing in \textsc{mesa} the \citet{GoughTayler1966} theory for convection in the presence of magnetic fields, which accounts for the reduced energy transfer at pressures $P\lesssim B^2/(8\upi)$, in agreement with \citet{Tremblay2015}. This alternative approach enabled us to compute self-consistent temperature profiles and to identify the modified boundary of the convection zone -- beyond which mixing is truly inhibited. 

Quantitatively, for the observed magnetic double-faced white dwarfs, we obtain a revised estimate of $B_{\rm crit}^{\rm He}(M,T_{\rm eff})\sim10^5$ G (top panel of Fig. \ref{fig:Grad}), orders of magnitude lower than our initial estimate. For two out of the four stars in our sample, the measured $B\gg B_{\rm crit}^{\rm H/He}$, probably ruling out the magnetic inhibition of convection as an explanation for their double-faced nature. For the remaining two stars, $B\gtrsim B_{\rm crit}^{\rm He}$ by about an order of magnitude, such that regions with a weaker magnetic field or a favourable orientation may possibly account for the surface inhomogeneity. 

However, given our mixed results, other scenarios where the magnetic field plays a different role -- if at all -- should be considered as well. For example, \citet{BedardTremblay2025} applied various mechanisms, such as magnetically reduced diffusion \citep{MichaudFontaine1982}, semi-convection, and lateral diffusion, to the prototypical double-faced white dwarf \citep{Caiazzo2023}, which does not currently have a magnetic field measurement. A similar quantitative analysis of the double-faced white dwarfs with measured magnetic fields could shed more light on this class.
 
While we tested the sensitivity of our results to the mixing-length parameter $\alpha$, other uncertainties remain. Specifically, convective mixing and dilution may depend on the unknown thickness of the superficial hydrogen atmosphere, and on the exact fractions of hydrogen and helium in the envelope \citep{Rolland2018,Rolland2020,Cunningham2020}. In other words, it is not clear whether the double-faced white dwarfs phenomenon can be reduced to a simple $B\sim B_{\rm crit}$ criterion. Specifically, $B_{\rm crit}$ may represent a lower threshold, with different outcomes possible for $B>B_{\rm crit}$. Another source of uncertainty is the treatment of convective overshoot, which plays an important role in the interplay between the hydrogen and helium layers \citep{Kupka2018,Cunningham2019,Cunningham2020,Bedard2024}.
A more comprehensive study that focuses on these uncertainties might bring the theory closer to the observations (or, alternatively, rule it out convincingly). 
Accurate fits to phase-resolved spectra of double-faced white dwarfs may also constrain some of the uncertainties \citep{BedardTremblay2025}. Another path forward is to increase the sample of observed double-faced white dwarfs. With more stars populating our figures, it could be easier to identify the potential role of magnetism in shaping this intriguing class, and specifically to distinguish between the roles of hydrogen and helium convection. 

In summary, this work -- with its inconclusive results -- motivates a more detailed analysis of the possible effects of magnetism on diffusion and mixing in double-faced white dwarfs. We propose to explicitly compute the envelope's structure, including the microscopic and macroscopic (i.e. due to convection) diffusion coefficients \citep[see][who compared them]{Cunningham2019} as a function of $B$, given each star's specific observational constraints \citep[see][]{BedardTremblay2025}. Such future work may identify the true role that magnetism plays in this class.



\section*{Acknowledgements}

I thank Adam Moss, Antoine B\'edard, Elad Shiftan, and especially the reviewer Pier-Emmanuel Tremblay for helpful discussions and comments that significantly improved the paper. 
This research was partially supported by the United States-Israel Binational Science Foundation (BSF; grant no. 2022175), the German-Israeli Foundation for Scientific Research and Development (GIF; grant no. I-1567-303.5-2024), the Israel Science Foundation (ISF; grant nos 1600/24 and 1965/24), and the Ministry of Innovation, Science, and Technology (grant no. 1001572596).   
 

\section*{Data Availability}

The data underlying this article will be shared on reasonable request to the corresponding author.



\bibliographystyle{mnras}
\input{double.bbl}




\bsp	
\label{lastpage}
\end{document}